\documentclass[conference]{IEEEtran}
\IEEEoverridecommandlockouts

\usepackage{cite}
\usepackage{amsmath,amssymb,amsfonts}
\usepackage{algorithmic}
\usepackage{graphicx}
\usepackage{textcomp}
\usepackage{xcolor}
\usepackage{color}
\usepackage{booktabs}
\usepackage{multirow}
\usepackage{threeparttable}
\usepackage{hyperref}
\hypersetup{hidelinks,
	colorlinks=true,
	allcolors=black,
	pdfstartview=Fit,
	breaklinks=true}
\def\BibTeX{{\rm B\kern-.05em{\sc i\kern-.025em b}\kern-.08em
    T\kern-.1667em\lower.7ex\hbox{E}\kern-.125emX}}
\begin{document}

\title{
WMCodec: End-to-End Neural Speech Codec with Deep Watermarking for Authenticity Verification
}


\makeatletter
\newcommand{\linebreakand}{%
  \end{@IEEEauthorhalign}
  \hfill\mbox{}\par
  \mbox{}\hfill\begin{@IEEEauthorhalign}
}
\makeatother

\author{\IEEEauthorblockN{1\textsuperscript{st} Junzuo Zhou}
\IEEEauthorblockA{\textit{Institute of Automation} \\
\textit{Chinese Academy of Sciences}\\
Beijing, China \\
zhoujunzuo2023@ia.ac.cn}
\and
\IEEEauthorblockN{2\textsuperscript{nd} Jiangyan Yi*}
\IEEEauthorblockA{\textit{Institute of Automation} \\
\textit{Chinese Academy of Sciences}\\
Beijing, China \\
jiangyan.yi@nlpr.ia.ac.cn}
\and
\IEEEauthorblockN{3\textsuperscript{rd} Yong Ren}
\IEEEauthorblockA{\textit{Institute of Automation} \\
\textit{Chinese Academy of Sciences}\\
Beijing, China \\
thurenyong@gmail.com}
\and
\IEEEauthorblockN{4\textsuperscript{th} Jianhua Tao}
\IEEEauthorblockA{\textit{Department of Automation} \\
\textit{Tsinghua University}\\
Beijing, China \\
jhtao@tsinghua.edu.cn
}
\linebreakand
\IEEEauthorblockN{5\textsuperscript{th} Tao Wang}
\IEEEauthorblockA{\textit{Institute of Automation} \\
\textit{Chinese Academy of Sciences}\\
Beijing, China \\
wangtao2018@ia.ac.cn}
\and
\IEEEauthorblockN{6\textsuperscript{th} Chu Yuan Zhang}
\IEEEauthorblockA{\textit{Department of Automation} \\
\textit{Tsinghua University}\\
Beijing, China \\
cy-z24@mails.tsinghua.edu.cn}
}

\maketitle
\footnotetext[1]{* Corresponding authors.}

\begin{abstract}

Recent advances in speech spoofing necessitate stronger verification mechanisms in neural speech codecs to ensure authenticity. Current methods embed numerical watermarks before compression and extract them from reconstructed speech for verification, but face limitations such as separate training processes for the watermark and codec, and insufficient cross-modal information integration, leading to reduced watermark imperceptibility, extraction accuracy, and capacity. To address these issues, we propose WMCodec, the first neural speech codec to jointly train compression-reconstruction and watermark embedding-extraction in an end-to-end manner, optimizing both imperceptibility and extractability of the watermark. Furthermore, We design an iterative Attention Imprint Unit (AIU) for deeper feature integration of watermark and speech, reducing the impact of quantization noise on the watermark. Experimental results show WMCodec outperforms AudioSeal with Encodec in most quality metrics for watermark imperceptibility and consistently exceeds both AudioSeal with Encodec and reinforced TraceableSpeech in extraction accuracy of watermark. At bandwidth of 6 kbps with a watermark capacity of 16 bps, WMCodec maintains over 99\% extraction accuracy under common attacks, demonstrating strong robustness. The code is available in \href{https://github.com/zjzser/WMCodec}{\color{blue}{WMCodec}}

\end{abstract}


\section{Introduction}

Speech codecs \cite{opusoldcodec0,huang2012oldcodec1,valin2009holdcodec2,bessette2002oldcodec4,hicsonmez2013oldcodec3} utilize an encoder-decoder pipeline to eliminate redundancies and produce a compact bitstream for effective compression and restoration. In recent years, neural speech codecs have exhibited remarkable performance \cite{{zeghidour2021soundstream,defossez2022encodec,ren2024ticodec,yang2023hifi}}. However, as awareness for Internet security increases, new security requirements have emerged for downstream applications of neural speech compression, including transmission and storage. With the growing sophistication of speech spoofing\cite{spoof1,spoof2,spoof3,spoof4}, it is vital for the decompression user in an asymmetric information exchange, such as a receiver of an unfamiliar bitstream, to verify that the recovered speech has been faithfully encoded and transmitted by the sender.

In response to this need, several approaches have been explored to embed numerical watermark as verification marks into speech during compression for protecting. Using specific algorithms to extract inaudible watermarks from reconstructed speech, these marks can serve as authentication markers for the decompression user to verify the authenticity of the codec process. Fig. \ref{fig:examplefig} illustrates a relevant example.

\begin{figure}[t]
\centering
  \centerline{\includegraphics[width=1.0\linewidth]{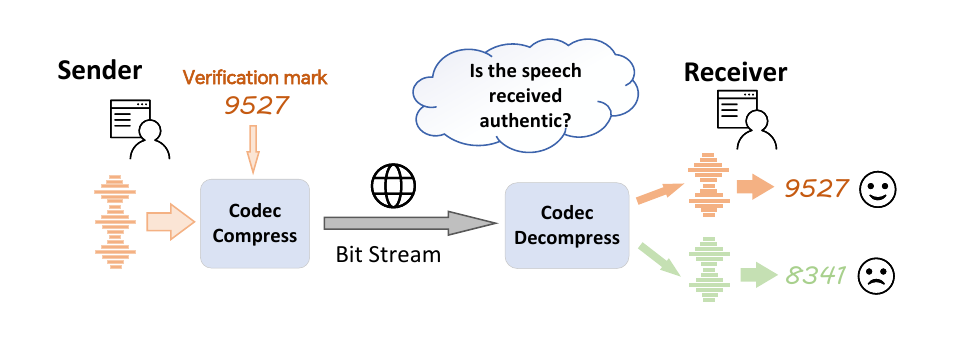}}
  \caption{
  Example of Watermark as Verification Marking for Codec Protection
  }
  \label{fig:examplefig}
\end{figure}

In recent years, deep-learning-based audio watermarking techniques have gradually developed, achieving notable performance. 
Chen et al. \cite{chen2023wavmark} proposed an robust audio watermarking framewok based on reversible networks. 
Liu et al. \cite{liu2023detecting} proposed a watermarking method for detecting voice cloning attacks. 
Zhou et al. \cite{zhou2024traceablespeech} utilized watermarking method for a speech synthesis model with proactively traceability. 
Roman et al. \cite{audioseal} surpasses the performance of previous work in watermark detection and localization.

However, to achieve verification-protected codecs, embedding watermarks into speech during compression through the above frameworks still has certain limitations.
Firstly, even advanced watermarking methods typically regard speech codecs as an untrained watermarking attack \cite{audioseal} or as an intermediate stage in a speech synthesis system \cite{zhou2024traceablespeech}. The separation optimization of the watermarking from the codec's quantization leads to error accumulation, which hinders the extractability and imperceptibility of the watermark. 
Secondly, the quantization noise in neural codecs necessitates a more adaptive and deeper fusion of watermark and speech features. In previous methods, the embedding of watermarks was based on simple modal fusion approaches, such as concatenation \cite{zhou2024traceablespeech,liu2023detecting} or addition \cite{audioseal}. The insufficient depth of this fusion limits the accuracy and capacity of watermark extraction.

\begin{figure*}[htbp]
\centering
  \centerline{\includegraphics[width=0.9\linewidth]{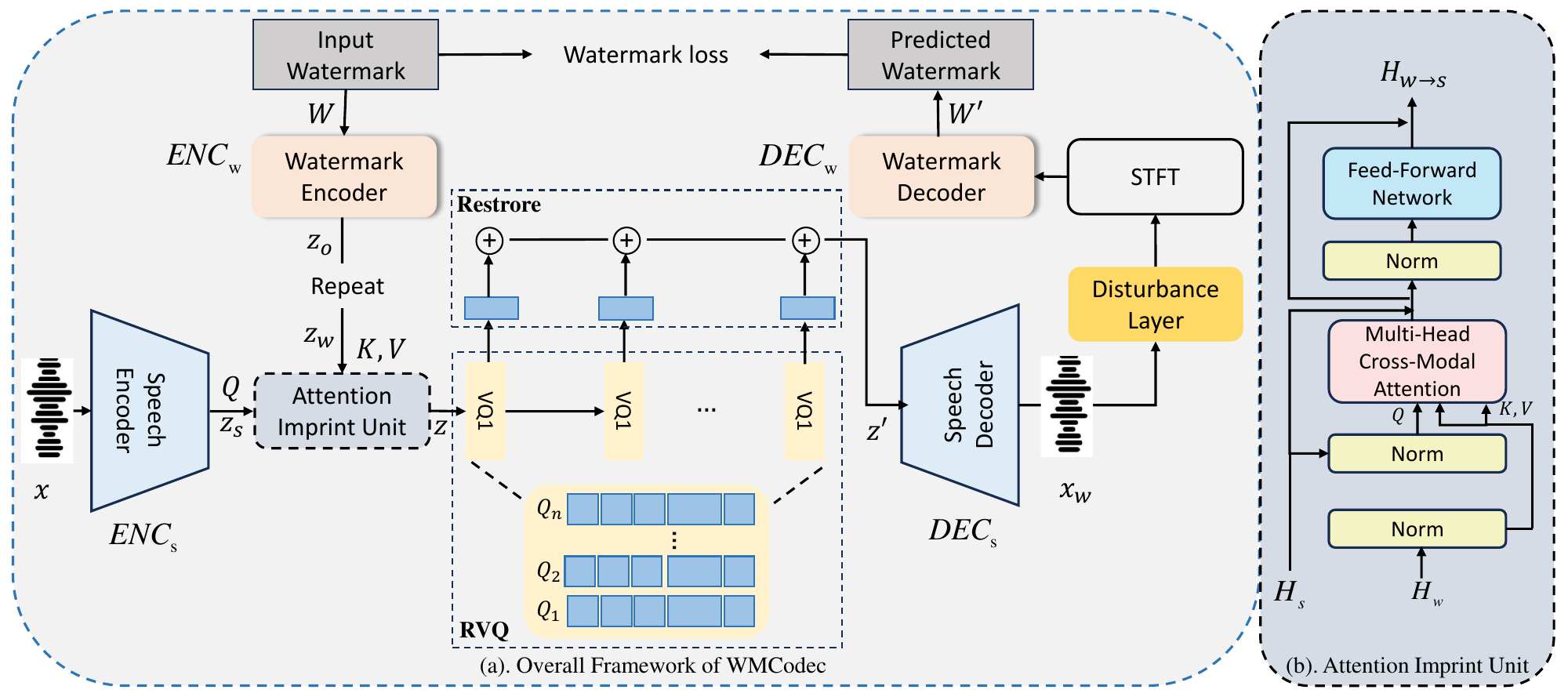}}
  \caption{
  Overview of the WMCodec framework.
  }
  \label{fig:framwork}
\end{figure*}

To address these issues, we propose a verification-protected neural speech codec named WMCodec.
Firstly, this model incorporates vector quantization between watermark embedding and extraction for end-to-end joint training. This mitigates the adverse effects of codec compression on the embedded watermark, thereby optimizing both the imperceptibility and extraction accuracy of the watermark. 
Secondly, we specifically design the iterative Attention Imprint Unit (AIU) for the codec. By leveraging cross-attention to induce a modal latent space for deep embedding \cite{sun2023emt}, it enables a more adaptive fusion of numerical watermark and speech information. AIU contributes to improving the accuracy and capacity of watermark extraction. 
Based on the two points outlined above, even after the codec process with limited bandwidth. the watermark in WMCodec can still be precisely extracted.

Our experiments on the LibriTTS \cite{zen2019libritts} dataset show that WMCodec achieves superior scores in both objective (STOI, PESQ) and subjective (MOS) metrics for watermark imperceptibility at bitrates of 3, 6, and 12 kbps, outperforming the strong baseline AudioSeal with Encodec. 
For watermark extractability, WMCodec consistently demonstrates a clear advantage over AudioSeal with Encodec and reinforced TraceableSpeech across various bandwidth. 
At bandwidth of 6 kbps with a watermark capacity of 16 bps, WMCodec maintains over 99\% extraction accuracy under common attacks, indicating strong robustness and practicality. 
Additionally, our ablation analysis further reveals that AIU significantly enhances extraction accuracy without compromising imperceptibility.

\section{Proposed Method}
\subsection{End to End Framework of WMCodec} 


To protect neural codecs through authenticity verification, the watermark should be embedded before the quantizer compression and accurately extracted after speech reconstruction. However, previous methods \cite{audioseal,zhou2024traceablespeech} did not integrate the codec into the watermarking-recovery training process. This separation of steps has made accurate watermark extraction challenging. WMCodec addresses this issue through end-to-end training, jointly optimizing watermark recovery and speech reconstruction, reducing the impact of codec-induced speech degradation on the watermarking. The customized framework is shown in Fig. \ref{fig:framwork}(a).

Let $x \in \mathbb{R}^{L_s}$ represents a speech signal, where $L_s$ denotes its length. The waveform $x$ is fed into the speech encoder $\mathbf{ENC}_{\mathrm{s}}$, undergoing multiple layers of downsampling to generate high-dimensional carrier features ${z}_{s}$:

\begin{equation}
{z}_{s} = \mathbf{ENC}_{\mathrm{s}}(x).
\end{equation}
Simultaneously, the $m$-digit base-$b$ numerical information $w$ is fed into the watermark encoder $\mathbf{ENC}{\mathrm{w}}$ \cite{zhou2024traceablespeech}. Specifically, each digit is mapped as embedding and uniformly concatenated. 
The entire watermark is then passed through two linear layers to obtain a single-frame feature ${z_o}$, which is temporally broadcast to align with ${z_s}$, forming the full-time watermark feature ${z_w}$. Through the iterative Attention Imprint Unit (AIU), ${z_w}$ is embedded into ${z_s}$, obtaining the combined feature $z$.

\begin{equation}
{z}_{w} = \mathbf{Repeat}(\mathbf{ENC}_{\mathrm{w}}(w)).
\end{equation}
\begin{equation}
{z} =  \mathbf{AIU}(z_s,z_w)
\end{equation}
where $z_s,z \in \mathbb{R}^{B \times T \times D_s}$, $z_w \in \mathbb{R}^{B \times T \times D_w}$. Afterward, $z$ is compressed via the residual vector quantizer $\mathbf{RVQ}$  \cite{zeghidour2021soundstream} and restored to $z'$, which is then decoded by speech decoder $\mathbf{DEC}{\mathrm{s}}$ to reconstruct the watermarked speech $x_w$.
\begin{equation}
z' = \mathbf{Restore}(\mathbf{RVQ}(z)).
\end{equation}
\begin{equation}
x_w = \mathbf{DEC}_{\mathrm{s}}(z').
\end{equation}

After the disturbance layer \cite{zhou2024traceablespeech}, during training, watermark decoder $\mathbf{DEC}{\mathrm{w}}$ processes the Mel-spectrogram of $x_w$ to extract the watermark $w'$. The decoder uses a high-dimensional vector decoded by ResNet \cite{zeinali2019but,wang2023wespeaker} , along with digit predictions generated by linear layers.
\begin{equation}
w' = \mathbf{DEC}_{\mathrm{w}}(\mathbf{Mel}(x_w)).
\end{equation}

\subsection{Attention Imprint Unit}\label{AA}
Previous methods for embedding numerical watermarks in speech often relied on simple multimodal fusion techniques, such as addition or concatenation of latent-dimensional features. 
However, due to the distortion arising from the compression-reconstruction, 
the WMCodec necessitates greater percolation of the numerical watermark features to facilitate the recovery of the watermark from carrier.
Therefore, in this work, an asymmetric cross-modal attention mechanism is leveraged to model inherent correlations between elements in two feature sequences, achieving the multimodal fusion.

\begin{table*}[t]
    \centering
    \caption{The main subjective and objective measures of our method and the baseline model's speech quality. }
        \begin{threeparttable}
            
        \begin{tabular}{l|c|c|cccc} 
            \toprule
            Model& N\_C (Bandwidth) & Capacity (bps) & PESQ $\uparrow$ & STOI $\uparrow$ & VISQOL $\uparrow$ & MOS $\uparrow$\\
            \midrule
            Audioseal + Encodec & \multirow{3}{*}{4 (3kbps)} & 16 & 2.134 & 0.894 & 3.643 & 2.908 $\pm$ 0.24\\ 
            WMCodec 4@16 (Ours) & & 16 & 2.606 & 0.898 & 3.549 & 4.152 $\pm$ 0.20\\
            WMCodec 4@10 (Ours) & & 13.3 & \textbf{2.848} & \textbf{0.912} & \textbf{3.758} & \textbf{4.458 $\pm$ 0.16} \\ 
            \midrule
            Audioseal + Encodec & \multirow{4}{*}{8 (6kbps)} & 16 & 2.569 & 0.931 & 3.861 & 3.819 $\pm$ 0.20\\
            TraceableSpeech* 4@16 & & 16 & 3.206 &0.933 & 4.014 & 4.458 $\pm$ 0.18\\ 
            WMCodec 4@16 (Ours) & & 16 & 3.187 & 0.936 & 4.009 & 4.434 $\pm$ 0.15 \\
            WMCodec 4@10 (Ours) & & 13.3 & \textbf{3.401} & \textbf{0.941} & \textbf{4.065} & \textbf{4.513 $\pm$ 0.12} \\ 
            \midrule
            Audioseal + Encodec & \multirow{3}{*}{16 (12kbps)} & 16 & 2.945 & 0.951 & 3.998 & 4.208 $\pm$ 0.19 \\
            WMCodec 4@16 (Ours) & & 16 & 3.558 & 0.953 & 4.138 & 4.535 $\pm$ 0.12 \\
            WMCodec 4@10 (Ours) & & 13.3 & \textbf{3.707} & \textbf{0.955} & \textbf{4.204} & \textbf{4.638 $\pm$ 0.13}\\
            \bottomrule
        \end{tabular}
    \label{tbl:1}

    \begin{tablenotes}    
        \footnotesize               
        \item[1] TraceableSpeech* denotes the modified TraceableSpeech in Section \ref{Baseline}
    
    \end{tablenotes} 
    \end{threeparttable}
\end{table*}

The architecture of Attention Imprint Unit(AIU) is shown in Fig. \ref{fig:framwork} (b).
In specific, considering $H_s$ and $H_w$ as the inputs, the calculation of AIU is as follows:
\begin{equation}
\begin{aligned}
H'_{w \to s} &= \text{MHCA}(\text{LN}(H_s), \text{LN}(H_w)) + H_s, \\
H_{w \to s} &= \text{FFN}(\text{LN}(H'_{w \to s})) + H'_{w \to s},
\end{aligned}
\end{equation}
where $\rightarrow$ denotes the information flow direction. 
$\text{LN}$ and $\text{FFN}$ represent layer normalization and the position-wise feed-forward network in Transformers, respectively. 
$\text{MHCA}$ refers to multi-head cross-modal attention, where Queries derive from the target modality $s$, Keys and Values derive from the source modality $w$, which facilitates latent adaptation from source to target.

The watermark embedding stage employs multiple AIU, allowing the speech feature to iteratively integrate information from the watermark in a progressive manner.

\subsection{Training Paradigm}

\noindent\textbf{Disturbance Layer:} During training, the disturbance layer simulates attacks on watermarked speech to ensure robustness under typical real-world operations. The simulation adopts an approach comparable to that of TraceableSpeech \cite{zhou2024traceablespeech}.

\noindent\textbf{Optimizing strategy}: 
During training, our optimization objectives include adversarial generator and discriminator losses, quantizer loss, and the cross-entropy loss \cite{yang2023hifi} between the extracted and original watermarks. This strategy minimizes distortion in both speech reconstruction and watermark recovery, similar to TraceableSpeech \cite{zhou2024traceablespeech}.

\begin{table*}[t]
    \centering
    \caption{Watermark extraction accuracy under various attacks. }
    \begin{threeparttable}
    \begin{tabular}{l|c|c|cccccccc|c}
    \toprule
    Model & N\_C (Bandwidth) & Capacity (bps) & Normal & RSP & Noise & SD & AR & EA & LP & Resplicing  & Average \\ 
    \midrule
    Audioseal + Encodec  & \multirow{3}{*}{4 (3kbps)} & 16 &0.540&0.514&0.523&0.522&0.528&0.526&0.532 & 0.514&  0.525\\
    WMCodec 4@16 (Ours) & &  16   
    & \textbf{0.762} & 0.765 & \textbf{0.754} & 0.765 & 0.768 & 0.764 & 0.750 & 0.711 & 0.755\\
     WMCodec 4@10 (Ours) & &  13.3 &0.760&\textbf{0.774}&0.738&\textbf{0.766}&\textbf{0.770}& \textbf{0.769} & \textbf{0.760} & \textbf{0.743} & \textbf{0.760}\\
    \cline{1-12}
    Audioseal + Encodec &  \multirow{4}{*}{8 (6kbps)} & 16 & 0.623 & 0.622 & 0.612 & 0.627 & 0.619 & 0.615 & 0.609 & 0.654 & 0.623\\
    TraceableSpeech* 4@16 & & 16 & 0.758 & 0.766 & 0.743 & 0.780 & 0.767 & 0.762 & 0.766 & 0.765 & 0.763\\
    WMCodec 4@16 (Ours) &  & 16 & 0.998 & 0.995 & 0.996 & 0.995 & 0.998 & \textbf{0.998} & \textbf{0.993} & 0.983& 0.995\\
    WMCodec 4@10 (Ours) & & 13.3 & \textbf{1.000} & \textbf{0.998} & \textbf{0.998} & \textbf{0.998} & \textbf{1.000} & 0.996 & 0.988 & \textbf{0.995} & \textbf{0.997}\\
    \cline{1-12}
    Audioseal + Encodec & \multirow{3}{*}{16 (12kbps)} & 16 &0.909&0.901&0.878&0.893&0.904&0.876&0.852& 0.815 &0.879\\
    WMCodec 4@16 (Ours) & & 16 &0.998&\textbf{1.000}&\textbf{1.000}&\textbf{1.000}&\textbf{1.000}&\textbf{0.998}&\textbf{1.000}& 0.967&0.995\\
    WMCodec 4@10 (Ours) & & 13.3 & \textbf{1.000}&0.998&0.996&\textbf{1.000}&0.995&0.997&\textbf{1.000}& \textbf{0.980} &\textbf{0.996}\\
    
    \bottomrule
    \end{tabular}
    \label{tbl:2}


    \begin{tablenotes}    
        \footnotesize               
        \item[1] The abbreviations in columns 5 to 10 correspond to common watermark attack operations, specifically: Resample(RSP), Random Noise(Noise), Sample Dropout(SD), Amplitude Reduction(AR), Echo Addition(EA), and Low-pass Filter(LP) \cite{zhou2024traceablespeech}.
        \item[2] Resplicing denotes a time-editing attack, where one-third of the utterance is randomly removed and the remaining segments are reconnected. \cite{zhou2024traceablespeech}.
        
    \end{tablenotes} 
    \end{threeparttable}
    
\end{table*}

\section{Experiments}
\subsection{Experiment Setup} \label{setup}
\subsubsection{Dataset} \label{dataset}
We utilize the LibriTTS dataset for training WMCodec. The LibriTTS \cite{zen2019libritts} corpus comprises 585 hours of English speech data, recorded at 24kHz, from 2,456 speakers. 
We randomly sampled 200 utterances from the LibriTTS test subsets, embedding watermarks into each utterance 10 times for evaluation experiment.

\subsubsection{Baseline} \label{Baseline}
AudioSeal \cite{audioseal} and TraceableSpeech \cite{zhou2024traceablespeech} are compared with our method, with each one being adapted for codec authenticity verification. 

\textbf{Audioseal:}
AudioSeal \cite{audioseal}  is a state-of-the-art audio watermarking method. As a general-purpose technique, it considers codec compression as one of the potential watermark attack operations. 
To implement a codec baseline using AudioSeal for verification mark protection, we incorporate a separately trained codec as an intermediate component in Audioseal. We use the open-source model weights "audioseal\_detector\_16bits" \footnote[2]{https://huggingface.co/facebook/audioseal} for inference. Correspondingly, the codec utilizes "encodec\_24khz"\footnote[3]{https://huggingface.co/facebook/encodec\_24khz} \cite{defossez2022encodec}, which is trained for 300 epochs, with one epoch being 2,000 updates.

\textbf{TraceableSpeech:}
TraceableSpeech \cite{zhou2024traceablespeech} is a speech synthesis system optimized for watermarking. While the codec serves as an intermediary, it operates solely before watermark embedding, preventing the watermark from acting as a verification mark for codec protection. Therefore, We now adjust the quantizer to operate between the watermark embedding and extraction, retraining the model using the original paper's parameters as a baseline. We denote the reinforced model as TraceableSpeech*, which is also a end-to-end model.


\subsubsection{Training details}
We train a 4-digit base-16 model (4@16) and a 4-digit base-10 model (4@10), with available watermark capacities of 16 bps and 13.3 bps, respectively.
For watermark embeding, we configure 2 iterations of AIU with 8 attention heads. Both the speech feature $z_s$ and the watermark feature $z_w$ are set to a dimension of 512. We adopted a HifiCodec-like structure \footnote[4]{https://github.com/yangdongchao/AcademiCodec} for the quantizer with a group size of 1. 
The number of codebooks is configured as 4, 8, and 16 for codec bandwidths of 3 kbps, 6 kbps, and 12 kbps, respectively.
We use ResNet221 \cite{zeinali2019but} in the watermark extraction.
All models are trained with a batch size of 32 and updated for 150K steps.

\subsection{Experiment Results and Analysis}\label{experiment_results}
\subsubsection{Performance of Watermark imperceptibility}

The imperceptibility of the watermark embedded in WMCodec can be evaluated by the quality of the reconstructed speech containing the watermark.
The original speech serves as the ground truth, and objective quality metrics such as PESQ \cite{rix2001pesq}, STOI\cite{taal2010stoi}, and VISQOL \cite{chinen2020visqol} are calculated between it and the reconstructed speech. For the subjective metric, we invited 7 skilled English users to rate the speech quality and reported the scores with a 95\% confidence interval. As shown in TABLE \ref{tbl:1}, under the same bandwidth and watermark capacity, WMCodec outperforms AudioSeal, which combines the pre-trained Encodec model, across most metrics. 
At bandwidths of 12kbps, WMCodec surpasses baseline in all quality metrics. 
Furthermore, compared to the modified TraceableSpeech, our method also demonstrates partial advantages in STOI.
As the watermark capacity increases, its imperceptibility progressively diminishes, which aligns with general trends and demonstrates the completeness of the experiment.

\subsubsection{Performance of Watermark Extraction Accuracy and Robustness}
For \textbf{watermark extraction accuracy}, both our method and the baseline are evaluated by counting the number of correctly recovered digits from the extracted watermark. As illustrated in TABLE \ref{tbl:2}, under the same watermark capacity, our model outperforms all baselines across various bitrates. This advantage is particularly pronounced at low and medium bitrates, where distortion caused by speech compression and reconstruction is more severe.
Moreover, at medium and high bitrates, the accuracy of our method is nearly 1.000, indicating that WMCodec can effectively utilize the watermark for authenticity verification.

For the \textbf{robustness} of watermark, as shown in TABLE \ref{tbl:2}, the extraction accuracy remains comparable to normal speech under various common watermark attacks. Furthermore, even under time-editing attacks like resplicing, WMCodec still outperforms other baselines. The experimental results demonstrate the strong robustness of WMCodec.

\subsubsection{Ablation Analysis}
Considering the modifications and retraining of TraceableSpeech in Section \ref{Baseline}, the comparison with this baseline can be regarded as an ablation analysis of the watermark embedding module.
The results for TraceableSpeech* in TABLE \ref{tbl:1} and \ref{tbl:2} indicate that the AIU significantly improves extraction accuracy over direct multimodal feature concatenation, while preserving comparable watermark imperceptibility.

\section{Conclusion}
This work proposes WMCodec, a model that end-to-end jointly optimizes the watermarking mechanism and neural codec process, thereby realizing the protection mechanism enables verification mark embedding before compression and extraction afterward.
This method also apply the iterative Attention Imprint Unit tailored of codec, which achieved deeper cross-modal integration for speech and numerical watermark. Experimental results demonstrate that WMCodec performs superiorly in most speech quality metrics, such as PESQ. In addition, this work 
perform better extraction accuracy than perivous method in most bandwidth and capacity scenarios. In the future, we aim to develop a codec for verification mark protection that achieves higher watermark capacity under lower bandwidth conditions.

\vfill\pagebreak

\bibliographystyle{IEEEtran}
\bibliography{refs}

\end{document}